\documentclass{ifacconf}

\makeatletter

\renewcommand\subsubsection{\@startsection{subsubsection}{3}{\z@}%
  {0.2\@bls \@plus .1\@bls}%
  {-0.5em}
  {\normalsize\itshape}}

\makeatother
\usepackage{graphicx}     
\usepackage{natbib}        
\usepackage{amssymb,amsfonts}
\usepackage{amsmath}
\DeclareMathOperator*{\argmin}{arg\,min}
\usepackage{textcomp}
\usepackage{xcolor}
\usepackage[short,nocomma,c2]{optidef}  
\usepackage{subcaption}
\usepackage{algorithm}
\usepackage{algorithmicx}
\usepackage{algpseudocode}
\usepackage{standalone}
\usepackage{tikz}
\usepackage{svg}
\usepackage{tabularx} 

\counterwithin*{section}{part}
\makeatletter
\let\old@ssect\@ssect
\makeatother
\newcommand{\mytitle}{Safe model-based Reinforcement Learning via Model Predictive Control and Control Barrier Functions}
\newcommand{\myauthor}{Kerim Dzhumageldyev}
\usepackage{hyperref}
\hypersetup{
    hidelinks,
    pdfauthor=\myauthor,
    pdfcreator=\myauthor,
    pdftitle=\mytitle,
}
\makeatletter
\def\@ssect#1#2#3#4#5#6{%
  \NR@gettitle{#6}
  \old@ssect{#1}{#2}{#3}{#4}{#5}{#6}
}
\makeatother

\algrenewcommand\algorithmicrequire{\textbf{Input:}}
\algrenewcommand\algorithmicensure{\textbf{Output:}}

\begin{document}
\begin{frontmatter}

\title{\mytitle} 

\author{\myauthor} 
\author{ Filippo Airaldi } 
\author{ Azita Dabiri}

\address{Delft Center for Systems and Control, Delft University of Technology, Mekelweg 2, 2628 CD Delft, The Netherlands \\ 
Emails: kerim.dzhum@gmail.com, \{f.airaldi, a.dabiri\}@tudelft.nl}

\begin{abstract}                
Optimal control strategies are often combined with safety certificates to ensure both performance and safety in safety-critical systems. A prominent example is combining Model Predictive Control (MPC) with Control Barrier Functions (CBF). Yet, efficient tuning of MPC parameters and choosing an appropriate class $\mathcal{K}$ function in the CBF is challenging and problem dependent. This paper introduces a safe model-based Reinforcement Learning (RL) framework where a parametric MPC controller incorporates a CBF constraint with a parameterized class $\mathcal{K}$ function and serves as a function approximator to learn improved safe control policies from data. Three variations of the framework are introduced, distinguished by the way the optimization problem is formulated and the class $\mathcal{K}$ function is parameterized, including neural architectures. Numerical experiments on a discrete double-integrator with static and dynamic obstacles demonstrate that the proposed methods improve performance while ensuring safety.
\end{abstract}

\begin{keyword}
Control Barrier Functions, Reinforcement Learning, Model Predictive Control
\end{keyword}

\end{frontmatter}

\section{Introduction}
Safety is a concern at the core of numerous automatic control problems. Over the years, various methodologies have been developed to ensure safety in controlled systems \citep{big_comparison}. Among these, in recent years Control Barrier Functions (CBFs) have become one of the most prominent tools. In particular, they enforce safety by ensuring that system trajectories remain within a prescribed safe set \citep{ames2019CBFTHEORYMAIN, xuames2017CBF}. More recently, CBFs have been combined with Model Predictive Control (MPC), which provides a predictive framework for optimizing control performance, while CBFs guarantee forward invariance of a safe set over the horizon \citep{mpccbfD}.

The effectiveness of this integration, however, hinges on a handful of design choices. Among others, weights in the MPC cost, the prediction horizon and the class $\mathcal{K}$ function in the CBF all shape the performance of the resulting controller. Poorly crafted components can shrink the feasible set, reducing performance or even render the problem infeasible. Notably, the choice of the class $\mathcal{K}$ function is particularly delicate, as it directly affects how conservative the safety constraint is and creates a trade-off between safety and performance.

To address the challenge of selecting a suitable class $\mathcal{K}$ function, several learning-based approaches have been proposed. One line of work models a generic class $\mathcal{K}$ function with a Neural Network (NN) trained jointly with a Reinforcement Learning (RL) policy \citep{learningclasskappa}. A Quadratic Program (QP) formulation enforces the learnable CBF condition ensuring safety, while RL alleviates the inherent myopic limitations of the QP. While QP-based methods enforce safety at each step and RL training helps curb myopic behavior, they do not propagate safety information or plan trajectories across a horizon, which can lead to infeasibility during training before feasible RL policies are found. By contrast, MPC enforces safety by optimizing over a prediction horizon rather than one-step ahead, thus being more likely to preserve feasibility more reliably during training \citep{nmpc}. Another approach introduces a learnable penalty term to modulate a user-specified class $\mathcal{K}$ function in the CBF condition \citep{barriernet}. Since the class $\mathcal{K}$ function itself is fixed, its choice remains arbitrary. In this approach, learning is performed through a differentiable QP with
supervised training, thus tying the performance to the quality of the nominal controller used for data generation. A third direction combines MPC–CBF with deep RL \citep{sabouni2024}. The RL policy is trained to output the parametrization for both the MPC controller and CBF constraints. This avoids differentiating through the optimization problem and reduces computation, but gradients are not propagated through MPC, making the method sample-inefficient. 

As an alternative, we propose using MPC as the function approximator instead of a NN as in \citet{sabouni2024}, and learning a parametrized class $\mathcal{K}$ function within the CBF constraint. Advantages of leveraging MPC as approximation include the explicit integration of prior expert knowledge (e.g., system dynamics, input and state constraints) into the policy, and the amenability of said approximation scheme to analysis and certification (e.g., stability, recursive feasibility). Compared to \citet{sabouni2024}, our approach is more sample efficient, since gradients are propagated through the MPC optimization problem rather than via black-box policy updates. Furthermore, while QP-based methods such as \citet{learningclasskappa,barriernet} remain limited by their myopic inability to plan coherent safe trajectories multiple steps ahead, the MPC scheme is able to yield policies that are optimal and guarantee safety across a prediction horizon. 

The contribution of this paper is to propose three different ways to parametrize the class $\mathcal{K}$ function in the context of MPC as function approximation for RL, each with its own distinct trade-offs:
\begin{itemize}
    \item \textit{Learnable Optimal-Decay CBF} (LOD-CBF) extends the optimal-decay framework from \citet{optdecay} by making decay parameters learnable, improving feasibility while maintaining safety guarantees.
    \item \textit{Neural Network CBF} (NN-CBF) replaces fixed decay rates with a feedforward NN that outputs state-dependent decay functions, removing horizon dependence and enabling richer mappings.
    \item \textit{Recurrent Neural Network CBF} (RNN-CBF) extends NN-CBF with a recurrent neural network (RNN) architecture that incorporates temporal context, improving performance in environments with time-varying constraints.
\end{itemize}

The paper is structured as follows. Section \ref{sec:theoretical_background} gives an an overview of MPC-based RL, along with discrete CBF, discrete exponetial CBF and optimal-decay CBF. This is used to motivate the methodology proposed in Section \ref{sec:methods}. The effectiveness of these methods is tested in Section \ref{sec:results} on two obstacle avoidance tasks. Lastly, Section \ref{sec:conclusion} concludes the paper and highlights future research directions. 
\section{Background}
In this section, we provide an overview of the concepts that will be used in the following sections.
\label{sec:theoretical_background}
\subsection{Control Barrier Functions}
Here we review discrete CBFs and their incorporation into MPC, followed by discrete exponential CBF and optimal-decay CBF, which form the basis of our approaches.

\subsubsection{Discrete CBFs and Exponential Variants.}

Consider a discrete-time system
\begin{equation}
    s_{t+1} = f(s_t,a_t),
    \label{eq:dctrlaffinesys}
\end{equation}
where $s_t \in \mathcal{S} \subseteq \mathbb{R}^{n_s}$ is the state and $a_t \in \mathcal{A} \subseteq \mathbb{R}^{n_a}$ is the action. The safe set is defined as the superlevel set of $h:\mathcal{S} \to \mathbb{R}$, i.e., $C = \{s \in \mathcal{S}  \mid h(s) \geq 0\}$. The discrete CBF condition ensures forward invariance of $C$ at time step $k$ by requiring that there exists an input $a_t$ in \eqref{eq:dctrlaffinesys} such that
\begin{equation}
h(s_{t+1}) - h(s_t) \geq -\alpha \left(h(s_t)\right), \quad \forall s_t \in \mathcal{S},
\label{eq:discrete_CBF}
\end{equation}
where $\alpha$ is a class $\mathcal{K}$ function satisfying $\alpha(r) \le r$, $\forall r > 0$ \citep{discretereg}. Embedding this condition into optimal control problems involves adding \eqref{eq:discrete_CBF} as an additional constraint. See \citet{mpccbfD} for MPC.

Discrete exponential CBFs extend nominal CBFs by selecting the class $\mathcal{K}$ function in \eqref{eq:discrete_CBF} as $\alpha(r)=\gamma r$, with $\gamma \in (0,1]$ controlling the decay rate, which imposes an exponential decay on the safety constraint over time \citep{eCBF, ames2019CBFTHEORYMAIN, discretereg}.

\subsubsection{Optimal-decay CBF.} \label{section:optd}
A limitation of the CBFs discussed above is that they may become infeasible under input constraints. Optimal-decay CBFs (OD-CBFs) address this limitation by replacing the constant decay rate $\gamma$ with the adjustable decision variable $\omega$, jointly optimized with the input \citep{optdecay} in the following QP
\begin{mini!}
    { {\scriptstyle u \in \mathcal{A},\ \omega \in (0,1]} }
    { 
        \frac{1}{2} \left\lvert u - \overline\pi(s_t) \right\rvert^2 
        + P_{\omega} ( \omega-\overline{\omega} )^2
    }{ \label{eq:optimaldecay} }{}
    \addConstraint{ h \left( f(s_t, u) \right) - h(s_{t}) }{ \ge -\omega h(s_t), }
\end{mini!}
where $\overline{\omega}$ is a reference decay rate, $P_\omega$ is a penalty weight and $\overline\pi : \mathcal{S} \to \mathcal{A}$ is a nominal controller. Note that in the limit $P_\omega \to \infty$ with $\overline{\omega}=1$, the constraint reduces to the nominal CBF. 

\subsection{MPC as function approximator in RL}
The objective of RL is to learn a deterministic policy $\pi : \mathcal{S} \to \mathcal{A}$ for the dynamic system \eqref{eq:dctrlaffinesys} that minimizes the sum of discounted costs
\begin{equation}
    J(\pi) \coloneqq \sum_{t=0}^T \zeta^t L \left( s_t, \pi(s_t) \right),
    \label{eq:discound_cumstagecost}
\end{equation}
where $L : \mathcal{S} \times \mathcal{A} \to \mathbb{R}$ is the stage cost, $\zeta \in (0, 1]$ is the discount factor and $T$ is the task length \citep{Sutton1998}. The optimal policy $\pi^\star$ is given then by:
\begin{equation}
    \pi^\star = \argmin_\pi J(\pi).
    \label{eq:opt_policy}
\end{equation}
In general, the true optimal policy $\pi^\star$ and corresponding value functions are difficult to compute exactly. One solution is to employ function approximation schemes to approximate these. Among others, such approximation schemes can be delivered by a parametrized MPC controller \citep{NMPCbasedRLGROS}, whereby the approximate parametrized value function $V_\theta$, estimating the cost-to-go from state $s$, is given by
\begin{mini!}
    { \scriptstyle X,U,\Sigma }
    {
        \lambda_\theta(x_0) 
        + \sum_{k=0}^{N-1}\zeta^k\left(\ell_\theta(x_k,u_k) + w^\top\sigma_k\right)
        \nonumber
    }
    { \label{eq:valfunc} }{ 
        \hspace{-4pt}  
        V_\theta(s) = 
    }
    \breakObjective{
        \hspace{25pt}  
        + \zeta^N \ell_\theta^\text{f}(x_N)
        \label{eq:valfunc_obj}
    }
    \addConstraint{x_{k+1} }{= f_\theta(x_k,u_k), \ }{ k=0,\ldots,N-1, }
    \addConstraint{ x_0 }{ = s_t, \label{eq:valfunc_first_constr} } 
    \addConstraint{ x_k }{ \in \mathcal{S}, }{ k=0,\ldots,N, }
    \addConstraint{ u_k }{ \in \mathcal{A}, \ \sigma_k \geq 0, }{ k=0,\ldots,N-1, \label{eq:valfunc_seclast_constr} }
    \addConstraint{ c_\theta(x_k,u_k) }{ \le \sigma_k, }{ k=0,\ldots,N-1. \label{eq:valfunc_last_constr} }
\end{mini!} 
Primal variables include the states $X = \left\{ x_k \right\}_{k=0}^N
$ and actions $U = \left\{ u_k \right\}_{k=0}^{N-1}$. The term $\lambda_\theta : \mathcal{S} \to \mathbb{R}$ represents the initial cost function, while $\ell_\theta : \mathcal{S} \times \mathcal{A} \to \mathbb{R}$ and $\ell_\theta^\text{f} : \mathcal{S} \to \mathbb{R}$ denote the stage and terminal ones respectively. The system dynamics are given by the parameterized model $f_\theta : \mathcal{S} \times \mathcal{A}\to S$. State–input constraints are given by $c_\theta : \mathcal{S} \times \mathcal{A} \to \mathbb{R}$, with slack variables $\Sigma = \left\{ \sigma_k \right\}_{k=0}^{N-1}$ ensuring feasibility. Their penalty $w$ controls the relaxation trade-off. The resulting policy applies the first optimal control input $\pi_\theta(s)=u_0^\star$. Note that the action-value function $Q_\theta(s,a)$ is defined analogously \citep{NMPCbasedRLGROS}:
\begin{equation}
    Q_\theta(s,a) = \min_{X,U,\Sigma} 
    \left\{ 
        \eqref{eq:valfunc_obj} \,:\, 
        \eqref{eq:valfunc_first_constr}\text{--}\eqref{eq:valfunc_last_constr},\ 
        u_0=a
    \right\}.
\end{equation}

The MPC-based RL framework can be coupled with different RL algorithms; in this work we focus on Q-learning \citep{watkins1989learning}. The objective is to minimize the Bellman residual error $\min_\theta \mathbb{E}\left[\|Q^\star(s, a) - Q_\theta(s, a)\|^2\right],$ in order to learn the optimal action-value function and subsequently recover the optimal policy from it. In practice, this amounts to driving the Temporal Difference (TD) error, defined as 
\begin{equation}
\tau_t = L(s_t, a_t) + \zeta V_\theta(s_{t+1}) - Q_\theta(s_s, a_s),
\label{eq:tderror}
\end{equation}
to zero with the parameter update step
\begin{equation}
\theta \gets \theta + \eta \tau_t \nabla_\theta Q_\theta(s_t, a_t),
\label{eq:qlearningparam}
\end{equation}
\noindent where $\eta > 0$ is the size of the update step. 
\section{Methodology}
\label{sec:methods}
This section proposes the three novel approaches for parametrizing the class $\mathcal{K}$ function of the CBF in MPC-based RL. As a general remark, note that the methods below parametrize also the MPC objective function to ensure a richer parametrization of the MPC controller is achieved.
\subsection{Learnable Optimal-decay CBF}

A key challenge in the OD-CBF framework is selecting suitable values for parameters $\overline{\omega}$ and $P_\omega$ introduced in \eqref{eq:optimaldecay}. As this selection is problem dependent, we propose to leverage RL to tune these quantities automatically. Consider a control task with $\mathcal{O}$ constraints (e.g., obstacles to avoid). We enforce each constraint $h_i$, $i=1,\ldots,\mathcal{O}$, via a separate CBF, which is in turn associated to learnable parameters $\overline{\omega}_{k,i,\theta}$ and $P_{\omega_{k,i},\theta}$, integrated into the cost function of MPC formulation as below.
\begin{mini!}
    { \scriptstyle X,U,\Omega,\Sigma }
    {
        G(X,U,\Sigma) 
        + \sum_{k=0}^{N-1}\sum_{i=1}^{\mathcal O}
        P_{\omega_{k,i},\theta} \left( \omega_{k,i} - \overline{\omega}_{k,i,\theta} \right)^2
        \label{eq:VOPTD_objective}
    }
    { \label{eq:VOPTD} }{}
    \addConstraint{ \eqref{eq:valfunc_first_constr}\text{--}\eqref{eq:valfunc_seclast_constr}, \label{eq:optd_first_const}}
    \addConstraint{x_{k+1} = f(x_k,u_k), \  k=0,\ldots,N-1,\label{eq:optd_second_const}}
    \addConstraint{
        0 < \omega_{k,i} \leq 1,
        \nonumber
    }
    \addConstraint{
        h_i(x_{k+1}) - (1-\omega_{k,i})\,h_i(x_k) \ge -\sigma_{k,i},
        \nonumber
    }
    \addConstraint{
        \hspace{40pt} k=0,\ldots,N-1, \ i=1,\ldots,\mathcal O.
    }
\end{mini!}
where
\begin{equation*}
    G(X,U,\Sigma)=\ell_{f,\theta}(x_N) + \sum_{k=0}^{N-1}\ell_{\theta}(x_k,u_k) + w_{\text{MPC}}\sum_{k=0}^{N-1}\sum_{i=1}^{\mathcal O} \sigma_{k,i},
\end{equation*} 
and the set $\Omega = \left\{ \omega_{k,i} \right\}_{k=0,i=1}^{N-1,\mathcal{O}}$ collects the decision variables that modulate the CBFs decay at each prediction step and constraint.  The slack variables $\Sigma = \left\{ \sigma_{k,i} \right\}_{k=0,i=1}^{N-1,\mathcal{O}}$ for each constraint $i=1,\ldots,\mathcal{O}$ are used to ensure feasibility by allowing temporary relaxation of the CBF constraints $i=1,\ldots,\mathcal{O}$, penalized in the cost through the weight $w_{\text{MPC}}$. This relaxation is necessary because, under input constraints, the CBF condition \eqref{eq:discrete_CBF} may not be feasible for all states $s$, and as such recursive feasibility cannot be guaranteed. While ensuring recursive feasibility is outside the scope of this paper, we remark that this relaxation mechanism is essential in preserving feasibility of the CBF constraints during training. With appropriate penalization via $w_\text{MPC}$, the RL agent can then learn to adhere to behaviors that are more likely to be safe during the implementation. 

By leveraging an RL algorithm, reference values $\bar\omega_{k,i,\theta}$ and penalty terms $P_{\omega_{k,i},\theta}$ are adjusted automatically to improve closed-loop performance, while the adaptive nature of the decay rates $\omega_{k,i}$ enhances feasibility and promotes minimal safety violations throughout the training process. However, note that the number of learnable parameters grows with both the prediction horizon and the number of constraints. Every constraint-step pair introduces a decay variable $\omega_{k,i}$ along with its own $\overline{\omega}_{k,i,\theta}$ and $P_{\omega_{k,i}, \theta}$. Longer horizons therefore increase not only the total number of constraints and optimal decision variables, but also the dimensionality of the parameter space to be tuned. Conversely, if the horizon is too short, the MPC scheme may lack sufficient flexibility to obtain a satisfactory policy. This strong dependence on the horizon is a principal limitation of the LOD-CBF method. 
\subsection{Neural Network CBF}
To address these limitations, we propose an alternative framework that reduces reliance on the horizon length and provides a richer parametrization of the class $\mathcal{K}$. It removes the CBF decay rates from the decision variables in \eqref{eq:VOPTD} and, instead employs a feedforward NN to directly model these decay rates denoted by $\gamma$ as a learnable function. 

At each prediction time step $k$, $k=0,\ldots,N-1$, the neural network takes the predicted state $x_k$, the CBF values $h_i(x_k)$, $i=1,\ldots,\mathcal{O}$, as well as the context variable $c_i(k)$ as input. The context variable $c_i(k)$ serves to encode any additional piece of information that can be beneficial to the decision making. The vector of output decay rates $\Gamma^{\text{NN}}_k = \begin{bmatrix} \gamma^{\text{NN}}_{k,1} & \dots & \gamma^{\text{NN}}_{k,\mathcal{O}} \end{bmatrix}^\top$ is then computed as follows:
\begin{subequations}
    \begin{align}
        z_k^{(0)} ={}& \begin{bmatrix}
            x_k & h_1(x_k) & \cdots & h_\mathcal{O}(x_k) & & c_1(k) & \cdots & c_\mathcal{O}(k)
        \end{bmatrix}^\top, \\
        z_k^{(j)} ={}& \mathrm{ReLU}\left(
            W_{\theta}^{(j)} z_k^{(j-1)} + b_{\theta}^{(j)}
        \right), \quad j = 1,\ldots,M, \\
        \Gamma_{k}^{\text{NN}} ={}& \mathrm{Sigmoid}\left(
            W_{\theta}^{(M+1)} z_k^{(M)} + b_{\theta}^{(M+1)}
        \right),
    \end{align}
\end{subequations}
where $\left\{ W_{\theta}^{(j)} \right\}_{j=1}^{M+1}$ and $\left\{ b_{\theta}^{(j)} \right\}_{j=1}^{M+1}$ denote the RL-learnable weight matrices and biases, respectively, with $M$ being the number of hidden layers. Applying the network over the entire prediction horizon provides the decay rates $\Gamma^{NN}_k$, $k=0,\ldots,N-1 $, that are substituted into the CBF constraints of the MPC formulation, constraint-wise and step-wise, yielding the following MPC scheme:

\begin{mini!}
    { \scriptstyle X, U, \Sigma }
    {G(X,U,\Sigma) \label{eq:NNCBF_objective} }
    { \label{eq:NNCBF} }{}
    \addConstraint{ \eqref{eq:optd_first_const}\text{--}\eqref{eq:optd_second_const}, }
    \addConstraint{
        h_i(x_{k+1}) - \left(1 - \gamma_{k,i}^{\text{NN}} \right) h_i(x_{k}) \ge -\sigma_{k,i}
        \nonumber
    },
    \addConstraint{
        \hspace{50pt} k=0,\ldots,N-1, \ i=1,\ldots,\mathcal O. \label{eq:NNCBF_CBFconst}
    }
\end{mini!}
Compared with the LOD-CBF, the class $\mathcal{K}$ functions in this formulation are decoupled from the MPC horizon and do not appear as decision variables. This property is particularly useful for shorter MPC horizons, where RL can still learn effective neural representations of the CBF decay rates, while the MPC controller maintains low computational overhead. Moreover, the NN can in general provide a richer and more scalable parametrization, enabling more expressive nonlinear mappings and finer trade-offs between safety and performance. However, reliance on a NN increases the number of trainable parameters and introduces non-trivial topological and hyperparameter choices. More notably, the proposed method lacks a mechanism to promote coherence of the predicted decay rates at consecutive time steps. Thus, these rates can potentially wildly oscillate as the dynamical system evolves, possibly comprising the performance and feasibility of the control policy from \eqref{eq:NNCBF}.

\subsection{Recurrent Neural Network CBF}
To further enhance the previous solution and circumvent the aforementioned issue, we propose to employ an Elman RNN \citep{elman} to similarly generate a vector of time-correlated decay rates across the prediction horizon. The RNN is defined by the following equations:
\begin{subequations}
    \begin{align}
        q_k^{(1)} ={}& \mathrm{ReLU}\left(
            W_{\theta}^{(1)} z_k^{(0)} + b_{\theta}^{(1)} + W_{q,\theta}^{(1)}q_{k-1}^{(1)}
        \right), \\
        \begin{split}
            q_k^{(j)} ={}& \mathrm{ReLU}\left(
                W_{\theta}^{(j)} q_k^{(j-1)} + b_{\theta}^{(j)} + W_{q, \theta}^{(j)}q_{k-1}^{(j)}
            \right), \\
            & \hspace{120pt} j = 2,\dots,M,
        \end{split} \\
        \Gamma_{k}^{\text{RNN}} ={}& \mathrm{Sigmoid}\left(
            W_{\theta}^{(M+1)} q_k^{(M)} + b_{\theta}^{(M+1)}
        \right),
    \end{align}
\end{subequations}
where $q_{k}^{(j)}$ denotes the hidden state of layer $j$ at time step $k$ and, similarly to the NN case, $\Gamma_k^{RNN} = \begin{bmatrix} \gamma^{\text{RNN}}_{k,1} & \dots & \gamma^{\text{RNN}}_{k,\mathcal{O}} \end{bmatrix}^\top$. The weight matrices and biases are again RL-learnable. The RNN is incorporated into the MPC value function in the same way as in \eqref{eq:NNCBF}; however, in \eqref{eq:NNCBF_CBFconst} $\gamma_{k,i}^{\text{NN}}$ is replaced by $\gamma_{k,i}^{\text{RNN}}$, as follows: 
\begin{mini!}
    { \scriptstyle X, U, \Sigma }
    { G(X,U,\Sigma)  \label{eq:RNNCBFobjective}}
    { \label{eq:RNNCBF} }{}
    \addConstraint{ \eqref{eq:optd_first_const}\text{--}\eqref{eq:optd_second_const}, }
    \addConstraint{
        h_i(x_{k+1}) - \left(1 - \gamma_{k,i}^{\text{RNN}} \right) h_i(x_{k}) \ge -\sigma_{k,i}
        \nonumber
    },
    \addConstraint{
        \hspace{50pt} k=0,\ldots,N-1, \ i=1,\ldots,\mathcal O.
    }
\end{mini!}
The key idea is that an RNN can learn to encode past information in its hidden states, retaining recent contextual information that a feedforward does not explicitly model. As a result, the RNN is able to handle time-varying constraints more effectively, since its structure is designed to exploit temporal relation. By leveraging their hidden state, RNNs can, under specific conditions, train more sample-efficiently than feedforward networks \citep{mousavihosseini2025RNN}. 

\subsection{Training Architecture} \label{subsec:training_architecture}

The overall training architecture is shown in Algorithm \ref{alg:rl_training}, where $n_{\text{ep}}$ denotes the total number of training episodes. Q-learning updates a parameter vector $\theta$ collecting all RL-learnable components, including the class $\mathcal{K}$ parameters $\{\overline{\omega}_{k,i,\theta}, P_{\omega_{k,i},\theta}\}$ in the LOD-CBF case, the feedforward weights and biases $\{W_{\theta}^{(j)}, b_{\theta}^{(j)}\}$ in the NN-CBF case, the additional recurrent weights $\{W_{q,\theta}^{(j)}\}$ in the RNN-CBF case, and the MPC objective parameters $\ell_{\theta}$ and $\ell_{f,\theta}$.
 
This RL training process is built around three MPC controllers. For each transition we solve $V_{\theta}(s)$ and $Q_{\theta}(s,a)$ based on \eqref{eq:VOPTD}, \eqref{eq:NNCBF} or \eqref{eq:RNNCBF}, depending on the choice of parametrization, which are used to form the TD error. To generate exploratory transitions, a behavioral policy based on the value function $V_\theta(s)$ (either one of \eqref{eq:VOPTD}, \eqref{eq:NNCBF} or \eqref{eq:RNNCBF}) is used. In particular, a perturbation term $\xi_t^\top u_0$ is added to objective of \eqref{eq:VOPTD_objective}, \eqref{eq:NNCBF_objective} or \eqref{eq:RNNCBFobjective}, where $\xi_t$ is sampled from a normal distribution every time the corresponding MPC problem is about to be solved. We refer to this modified controller as $\tilde{V}_{\theta}(s, \xi_t)$. The noise level decays over the course of training, allowing the policy to gradually shift from exploration to exploitation. 

An integral part of the training architecture is the RL stage cost $L(s,a)$, which encodes the return of the task at hand as per \eqref{eq:discound_cumstagecost}. The stage cost here is augmented to penalize the agent for the use of the slack variables as follows.
\begin{equation}
\tilde{L}(s_t,a_t, \Sigma_k^\star) = L\left(s_t,a_t\right) + w_{\tiny \text{RL}}\sum_{k=0}^{N-1}\sum_{i=1}^\mathcal{O} \sigma_{k,i}^\star 
\label{eq:stagecost_aug}.
\end{equation}
The second term in \eqref{eq:stagecost_aug} aggregates the optimal slack variables $\Sigma^\star$ from the optimal solution of the current $\tilde{V}_{{\theta}}(s_k, \xi_t)$. The weight $w_{RL}$ controls how strongly these violations are penalized during learning.

Gradients for the Q-learning update are first accumulated in a buffer and then averaged before updating the parameters. The averaged gradient is computed as
\begin{equation}
    \hat{g} = - \frac{1}{|\mathcal{B}|} \sum_{i=1}^{|\mathcal{B}|} \tau_i \nabla_{\theta} Q_{\theta}(s_i, a_i),
\end{equation}

\noindent where $|\mathcal{B}|$ denotes the buffer cardinality. In place of the plain gradient descent update \eqref{eq:qlearningparam}, we propose to leverage the Adam optimizer instead, which adaptively rescales learning rates by maintaining running estimates of first and second moments of the gradients, often leading to faster and more stable convergence \citep{kingma2015adam}. 

In case the space of the RL parametrization $\theta$ is bounded, lower and upper bounds can be enforce by projecting the update step onto the feasible bounds via a simple QP, as done in \citet{Rl_explanation}. In Algorithm \ref{alg:rl_training} we refer to this projected update step as $\mathrm{AdamQP}(\theta, \hat{g})$. 

\begin{algorithm}
    \caption{RL Training Loop}
    \label{alg:rl_training}
    \begin{algorithmic}[1]
    \For{episode $=1$ to $n_{\text{ep}}$}
        \State Reset RNN hidden states (if used)
        \For{$t = 0$ to $T$}
            \State \textbf{Exploration:} Solve $\tilde{V}_{{\theta}}(s_t,\xi_t)$ for $a_t$ and $\Sigma_t^\star$

            \State Observe stage cost $\tilde{L}(s_t,a_t, \Sigma_t^\star)$
            
            \State \textbf{System rollout:} Update $s_{t+1} \leftarrow f(s_t, a_t)$
            
            \State \textbf{Q-value:} Solve $Q_{\theta}(s_t,a_t)$ 
            
            \State \textbf{V-value:} Solve $V_{\theta}(s_{t+1})$ 
            
            \State \textbf{TD error:} Compute TD error \eqref{eq:tderror}
            
            \State \textbf{Compute Gradient:} $\nabla_{\theta}Q_{\theta}(s_t,a_t)$,
            \State Form $g_t \leftarrow -\tau_t \, \nabla_{\theta} Q_{\theta}(s_t,a_t)$ and store in $\mathcal{B}$
        \EndFor
        \If{buffer $\mathcal{B}$ is full}
            \State $\hat{g} \leftarrow \tfrac1{|\mathcal{B}|}\sum g_t$ and clear buffer $\mathcal{B}$
            \State Update $\theta\leftarrow\mathrm{AdamQP}(\theta, \hat{g})$
        \EndIf
        
        \State Decay exploration noise 
    \EndFor
    \end{algorithmic}
\end{algorithm}

\section{Results} \label{sec:results}

We test our proposed methodologies on two obstacle avoidance tasks. The first task evaluates the LOD-CBF and the NN-CBF on a simple example with one static obstacle. The second part considers a more complex environment with multiple moving obstacles, where the RNN-CBF and NN-CBF are compared to analyze their relative performance. Both set of experiments use the discrete LTI 2D double-integrator system used in \citet{mpccbfD}, with sampling time $\Delta t = 0.2 \text{s}$.

In all the results that follows, the control policy is provided by an MPC controller with 
\begin{equation}
\begin{aligned}
G(X,U,\Sigma) =\;& x_N^{T}F_{\theta}x_N 
+ \sum_{k=0}^{N-1}\left(x_k^{T}Q x_k + u_k^{T}R\,u_k\right) \\
& +\, w_{\text{MPC}}\sum_{k=0}^{N-1}\sum_{i=1}^{\mathcal O} \sigma_{k,i}.
\end{aligned}
\end{equation}

where $Q=10\mathbb{I}_{4\times4}$, $R=\mathbb{I}_{2\times2}$, and a learnable terminal weight $F_\theta$ initialized with $100\mathbb{I}_{4\times4}$, with zero off-diagonal entries and constrained to be positive definite. The system is subject to the boxing bounds $\mathcal{S} = \left\{ s \in \mathbb{R}^4 : \lVert s \rVert_\infty \leq 5 \right\}$, and $\mathcal{A} = \left\{ a \in \mathbb{R}^2 : \lVert a \rVert_\infty \leq 1 \right\}.$ All obstacles are modeled as $h(s) = (s_0 - x_o)^2 + (s_1 - y_o)^2 - r_o^2$, with $(x_o, y_o)$ the center and $r_o$ the radius of obstacle $o$. The RL stage cost is chosen as $L(s_t,a_t) = s_t^{T}Q s_t + a_t^{T}Ra_t $.

\subsection{Static Obstacle}

The first experiment considers the task of avoiding a single static obstacle. The system starts at $(-5,-5)$, and aims to reach the origin. The obstacle is centered at $(2, 2.25)$, with radius $1.5$. For both of the following algorithms, we opt for a myopic prediction horizon of one to start the learning from an extremely suboptimal policy. The parameters of this experiment are $w_{\text{MPC}}=20^6$ and $w_{\text{RL}}=10^3$. For LOD-CBF, we initialize $\overline{\omega}_{1,1,\theta}=1000$ and $P_{\omega_{1,1},\theta}=0.4$, while for NN-CBF we set the hidden layers of the network to $M=3$. 

\subsubsection{LOD-CBF.}

The initial LOD-CBF policy, illustrated in Fig. \ref{fig:optd_innit_policy}, is highly suboptimal, achieving a cumulative cost of around \textbf{21712}. Due to its myopic nature, the control policy fails to anticipate the obstacles and drives the system straight up to its boundary, after which it slides along it to maintain safety under the CBF condition. After training, the cumulative cost decreases to \textbf{7156}. As shown in Fig. \ref{fig:optd_final_policy}, the trajectory no longer moves directly toward the goal but instead deviates left to avoid the obstacle. The system also reaches the target in fewer steps due to higher velocities on average.

The improved behavior is also evident in the optimal $\omega^\star$ depicted in Fig. \ref{fig:optd_final_policy_omega}. A pronounced spike appears near the obstacle, reflecting the greater relaxation needed to maintain feasibility of the CBF constraint at higher velocity. These changes are directly influenced by the learned terminal cost parameters, shown in Fig. \ref{fig:optd_param_P}. The position weights ($F_{\theta_{1,1}}$ and $F_{\theta_{2,2}}$) dominate the velocity weights ($F_{\theta_{3,3}}$ and $F_{\theta_{4,4}}$), enabling higher velocities. Moreover, as $F_{\theta_{3,3}} \geq F_{\theta_{4,4}}$ at convergence, the control policy has learned to steer the system left, anticipating the obstacle. We remark that $\overline{\omega}_{1,1,\theta}$ and $P_{\omega_{1,1},\theta}$ did not change significantly.

\subsubsection{NN-CBF.}
\begin{figure}
    \centering
    \begin{subfigure}{0.48\columnwidth}
        \includegraphics[width=\columnwidth]{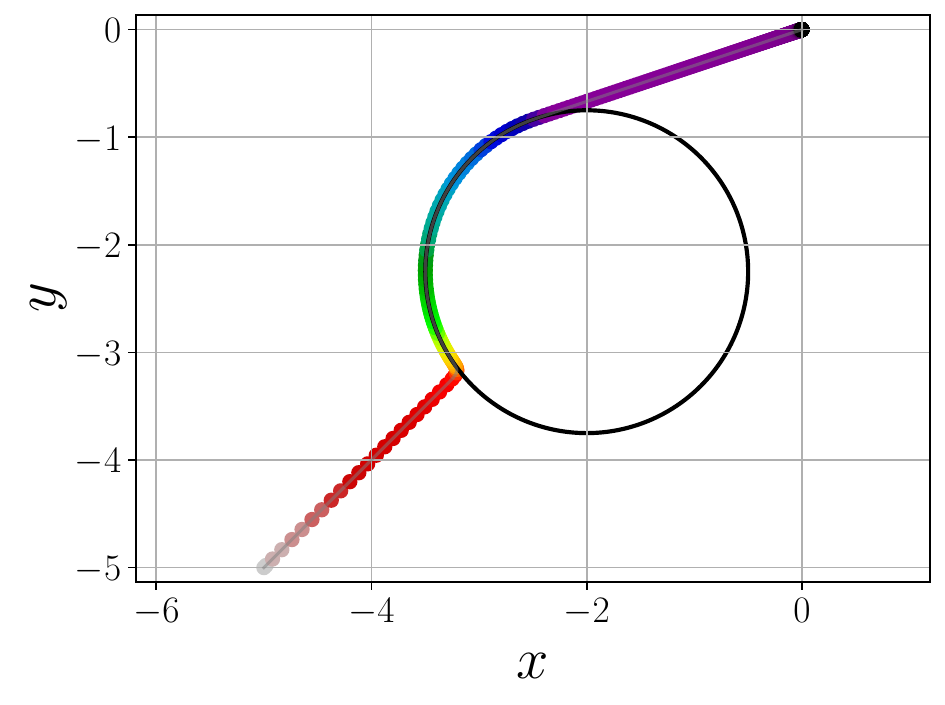}
        \captionsetup{aboveskip=0pt, belowskip=0pt} \caption{}
        \label{fig:optd_innit_policy}
    \end{subfigure}
    \begin{subfigure}{0.47\columnwidth}
        \includegraphics[width=\columnwidth]{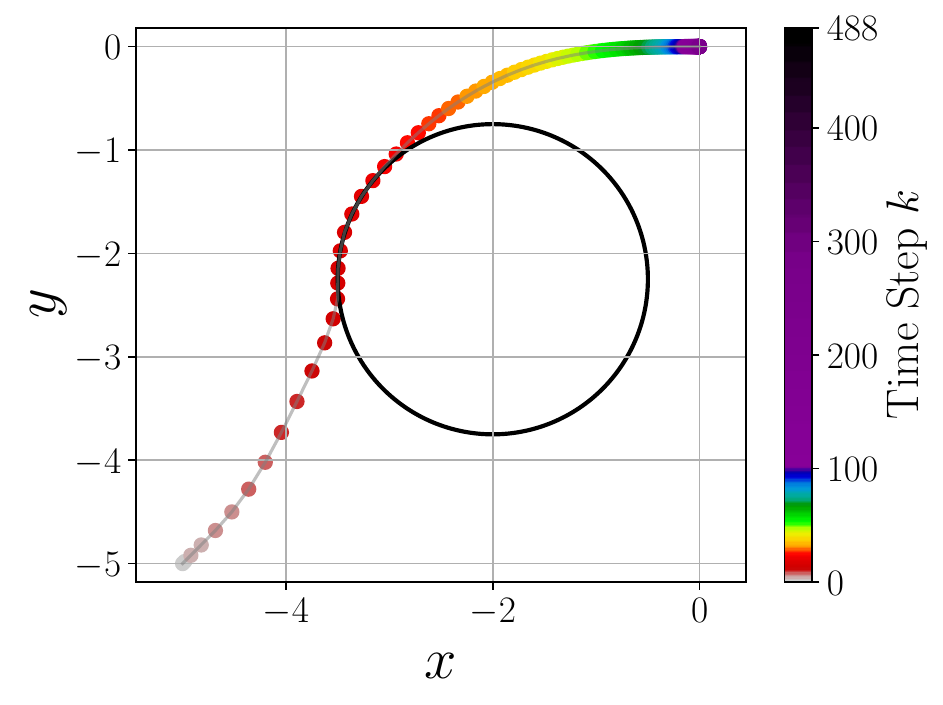}
        \captionsetup{aboveskip=0pt, belowskip=0pt} \caption{}
        \label{fig:optd_final_policy}
    \end{subfigure}
    \begin{subfigure}{0.48\columnwidth}
        \includegraphics[width=\columnwidth]{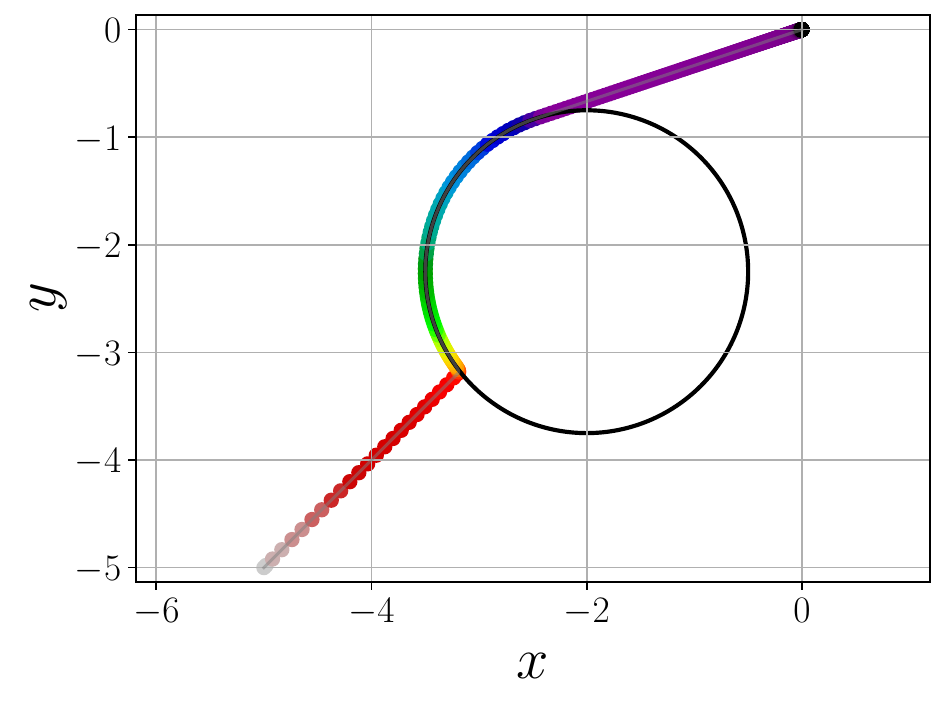}
        \captionsetup{aboveskip=0pt, belowskip=0pt} \caption{}
        \label{fig:sigmoidnn_innit_policy}
    \end{subfigure}
    \begin{subfigure}{0.47\columnwidth}
        \includegraphics[width=\linewidth]{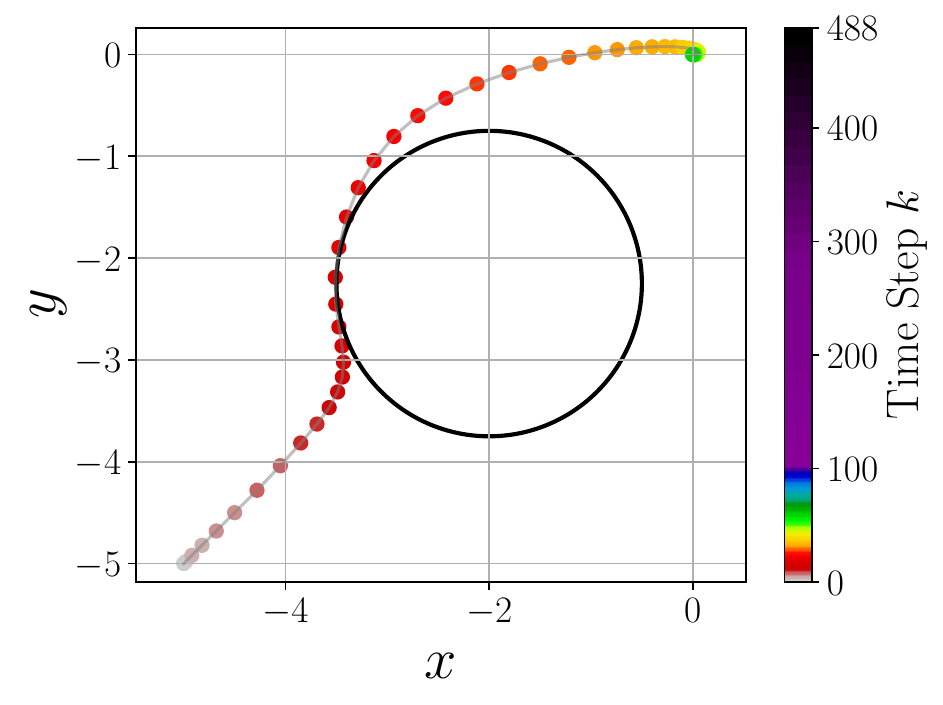}
        \captionsetup{aboveskip=0pt, belowskip=0pt} \caption{}
        \label{fig:sigmoidnn_final_policy}
    \end{subfigure}
    \begin{subfigure}{0.47\columnwidth}
        \includegraphics[width=\linewidth]{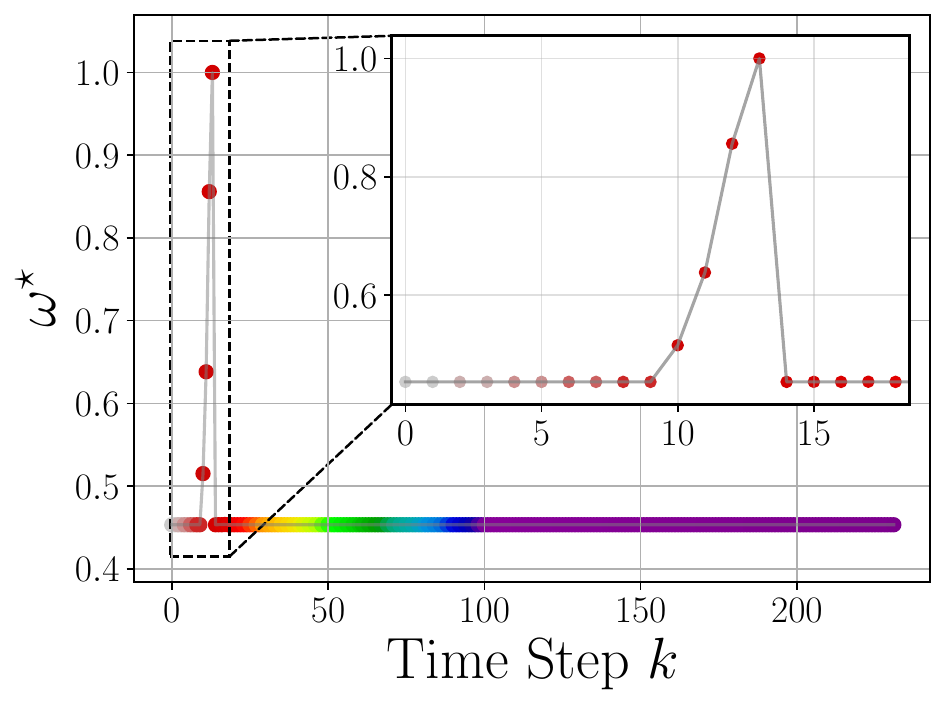}
        \captionsetup{aboveskip=0pt, belowskip=0pt} \caption{}
        \label{fig:optd_final_policy_omega}
    \end{subfigure}
    \begin{subfigure}{0.47\columnwidth}
        \includegraphics[width=\linewidth]{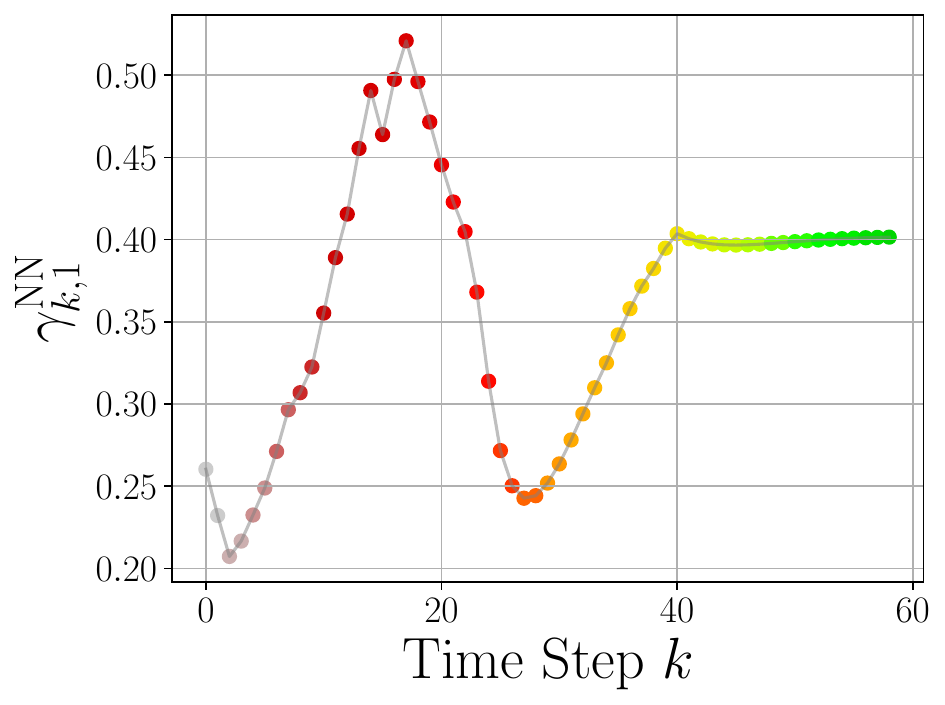}
        \captionsetup{aboveskip=0pt, belowskip=0pt} \caption{}
        \label{fig:sigmoidnn_final_policy_gamma}
    \end{subfigure}
    \begin{subfigure}{0.47\columnwidth}
        \includegraphics[width=\linewidth]{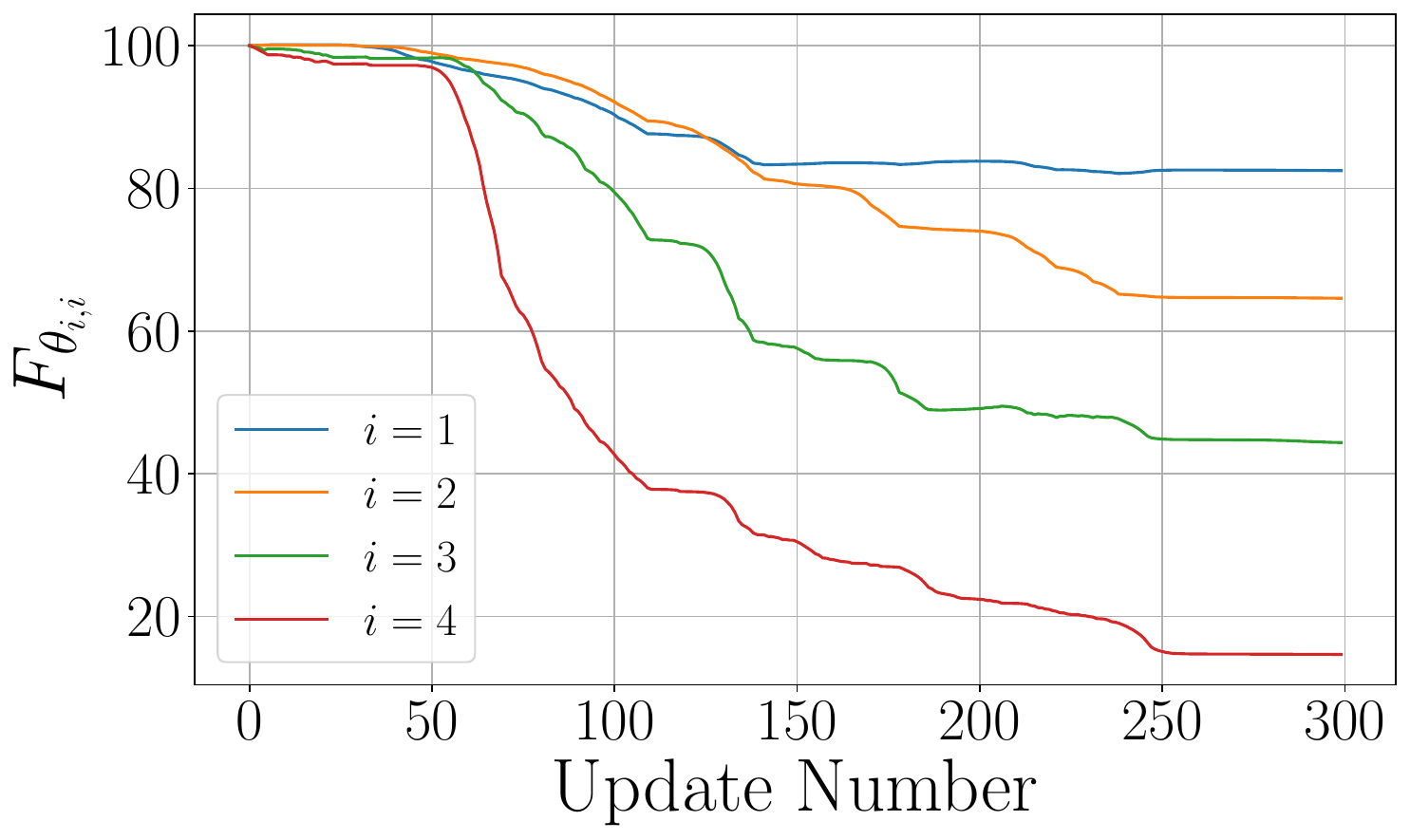}
        \captionsetup{aboveskip=0pt, belowskip=0pt} \caption{}
        \label{fig:optd_param_P} 
    \end{subfigure}
    \begin{subfigure}{0.47\columnwidth}
        \includegraphics[width=\linewidth]{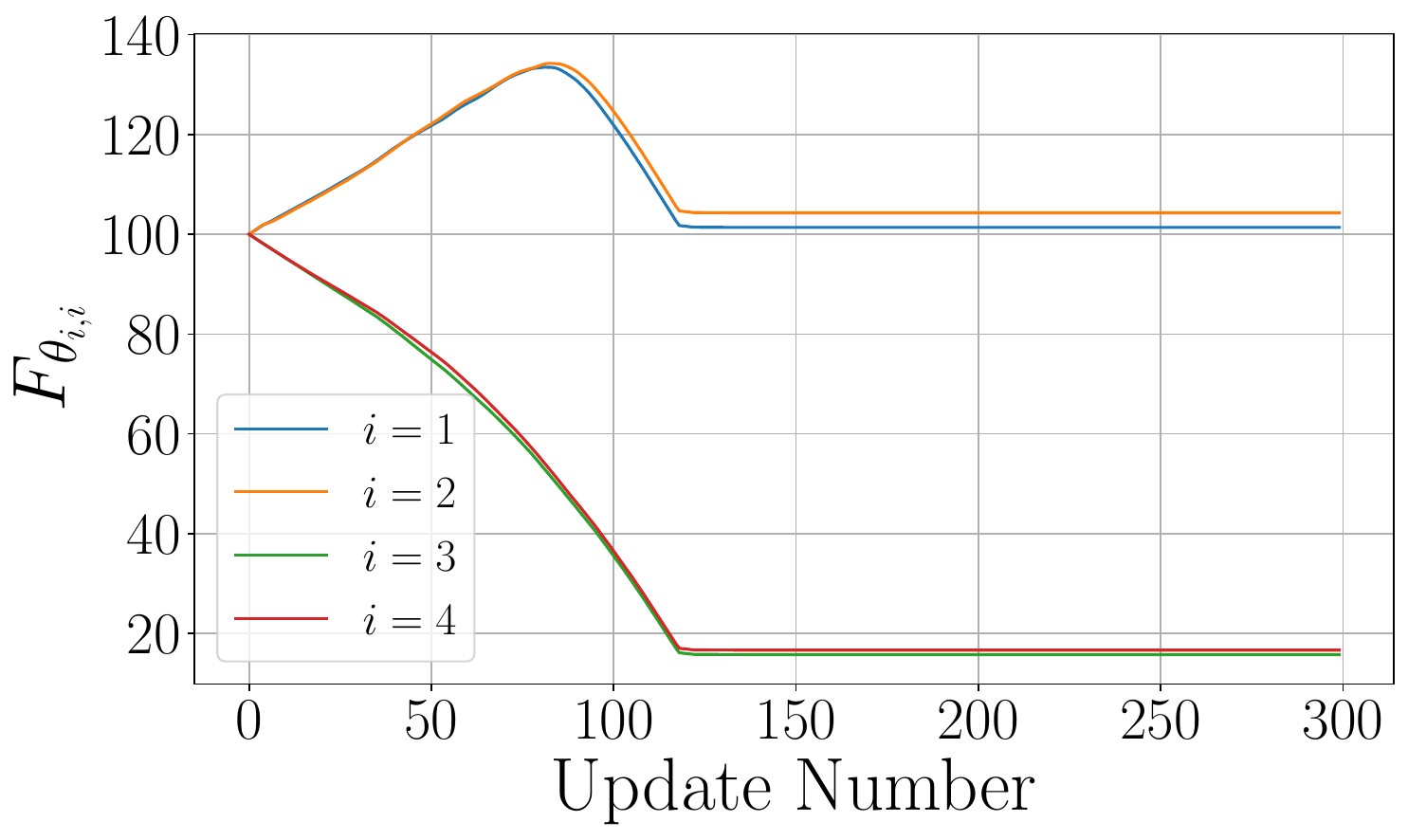}
        \captionsetup{aboveskip=0pt, belowskip=0pt} \caption{}
        \label{fig:sigmoidnn_param_P}
    \end{subfigure}
    \caption{
        Results for the static obstacle avoidance task.
        Trajectories, respectively at the start and end of training, for the LOD-CBF (a,b) and NN-CBF (c,d) approaches.
        Evolution at convergence of the LOD-CBF (e) optimal and NN-CBF (f) neural decay rates.
        Evolution of the terminal cost matrix elements for the LOD-CBF (g) and NN-CBF (h) approaches.
    }
\end{figure}

For the NN-CBF, the initial myopic policy is again suboptimal, with a cumulative cost of \textbf{21892} (Fig. \ref{fig:sigmoidnn_innit_policy}). Again, the RL training is effective in reducing the cost to \textbf{6627}, with the improved policy resulting in the trajectory shown in Fig. \ref{fig:sigmoidnn_final_policy}.

The improved behavior can be partially explained by the NN output decay rate $\gamma^\text{NN}_{k,1}$, shown in Fig. \ref{fig:sigmoidnn_final_policy_gamma}. At the start of the episode, $\gamma^\text{NN}_{k,1}$ decreases, preventing rapid acceleration, to then increase to enable faster progress toward the goal. Near the obstacle, the decay rate decreases again, introducing conservatism that facilitates a safe turn around the circle. After passing the obstacle, $\gamma^\text{NN}_{k,1}$ rises to permit faster motion again, before decreasing once more as the system approaches the target, preventing overshoot. The terminal cost parameters also influence the outcome. As shown in Fig. \ref{fig:sigmoidnn_param_P}, the position weights dominate the velocity weights, promoting faster motion, while the nearly equal velocity weights encourage forward rather than leftward steering. Nevertheless, the trajectory still turns left, indicating that the deviation is shaped not only by the terminal cost but also by the CBF constraint.

Comparing NN-CBF to LOD-CBF, it is evident that NN-CBF achieves better performance, which can be attributed to two factors. First, the NN can represent more complex class $\mathcal{K}$ functions, owing to its richer parametrization, enabling the CBF to encode more nuanced safety behavior. Moreover, the decay rate produced by the NN-CBF is purely state dependent, unlike the LOD-CBF whose decay rate is tied to the prediction horizon. As a result, the NN-CBF can learn steering behavior driven directly by the CBF, even when the terminal cost weights $x$ and $y$ are equal.

\subsection{Dynamic Obstacles}
In the second experiment, the system starts again in $(-5,5)$ and must navigate to the origin, this time avoiding one static and two moving obstacles. The two dynamic obstacles, with radii $0.7$, move horizontally at constant velocity, the first between coordinates $(-4,-1.5)$ and $(0,-1.5)$ and the second between $(-4,-3.3)$ and $(1,-3.3)$. The static obstacle is instead placed at $(-2,0)$ with radius $1$. The two dynamic obstacles hinder the system as it moves toward the origin, while the static obstacle discourages it from going around and forces the controller to find a trajectory that passes between the moving obstacles toward the goal. The increased difficulty of the task motivates using an MPC horizon of 6. The parameters of this experiment are $w_{\text{MPC}}=20^7$, $w_{\text{RL}}=10^5$, with $M=3$ for both the NN-CBF and the RNN-CBF. 

\subsubsection{NN-CBF.}

Fig. \ref{fig:NN_RNN_policy_snapshots} shows snapshots of the NN-CBF policy before and after training. Before training, the trajectory is unsafe, colliding with the first dynamic obstacle at $k=16$ and $k=17$, and the MPC predictions also fail to avoid future obstacle positions. After training, the trajectory slows earlier, remains outside the obstacle at $k=16$ to $k=18$, and maneuvers safely around it. The predicted trajectory also avoids further collisions with the dynamic obstacle, indicating improved planning and achieving a cumulative cost of \textbf{15194}.

\begin{figure}
    \centering
    \begin{subfigure}[b]{0.49\columnwidth}
        \includegraphics[width=\columnwidth]{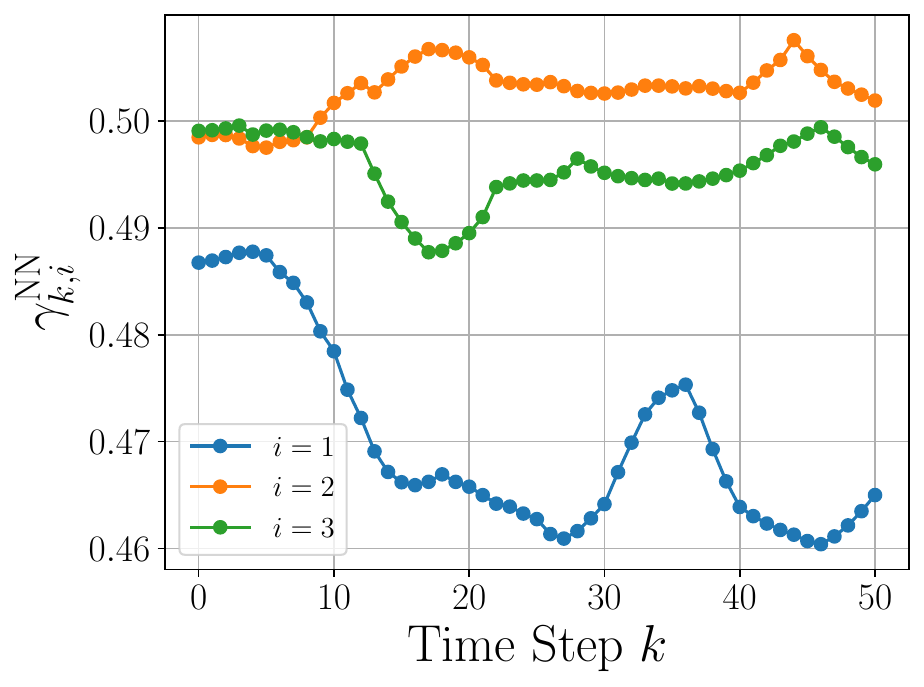}
        \caption{$\gamma^\text{NN}_{k,i}$ values of the NN}
        \label{fig:part2_aftertraining_gamma_NN}
    \end{subfigure}
    \begin{subfigure}[b]{0.49\columnwidth}
        \includegraphics[width=\columnwidth]{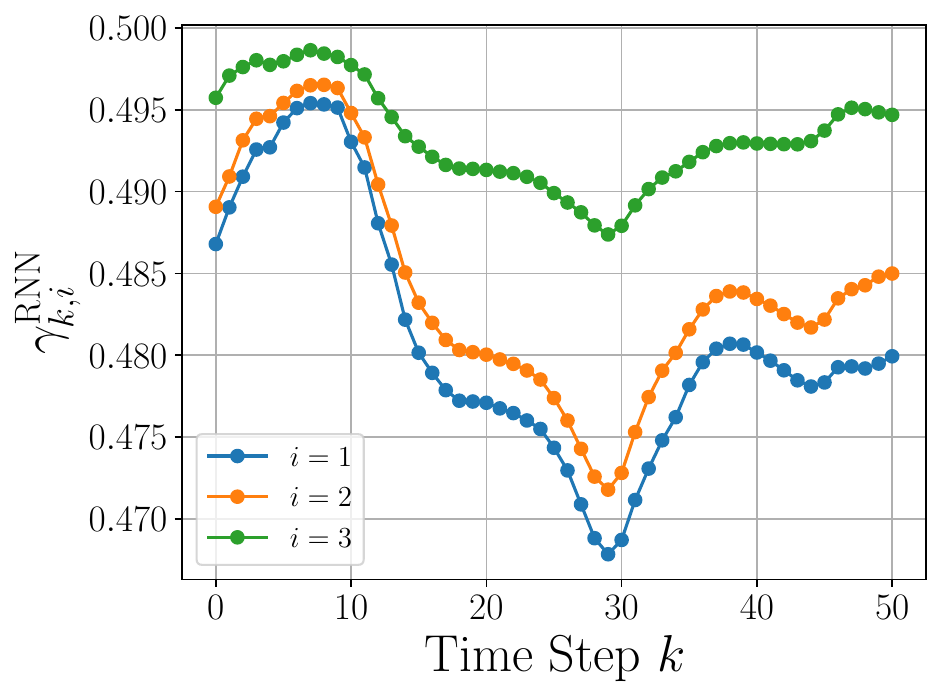}
        \caption{$\gamma^\text{RNN}_{k,i}$ values of the RNN}
        \label{fig:part2_aftertraining_gamma_RNN}
    \end{subfigure}
    \caption{NN and RNN outputs after training.}
\end{figure}

This improvement is also reflected in the NN output decay rates (Fig. \ref{fig:part2_aftertraining_gamma_NN}). The largest change occurs in $\gamma^\text{NN}_{k,1}$ for the first dynamic obstacle, whose constraint was previously violated. After training, $\gamma^\text{NN}_{k,1}$ decreases, causing the system to slow earlier when approaching the obstacle. 

\subsubsection{RNN-CBF.}
The training for the RNN-CBF approach follows a similar outcome to the previous case. However, the RNN-CBF policy appears to be less conservative, as the trajectory passes closer to the obstacle at $k=17$. This is further motivated by the cumulative cost achieved of \textbf{14026}, which is lower than the NN-CBF counterpart.

\begin{figure*}[h]  
  \centering
  \includegraphics[width=\textwidth]{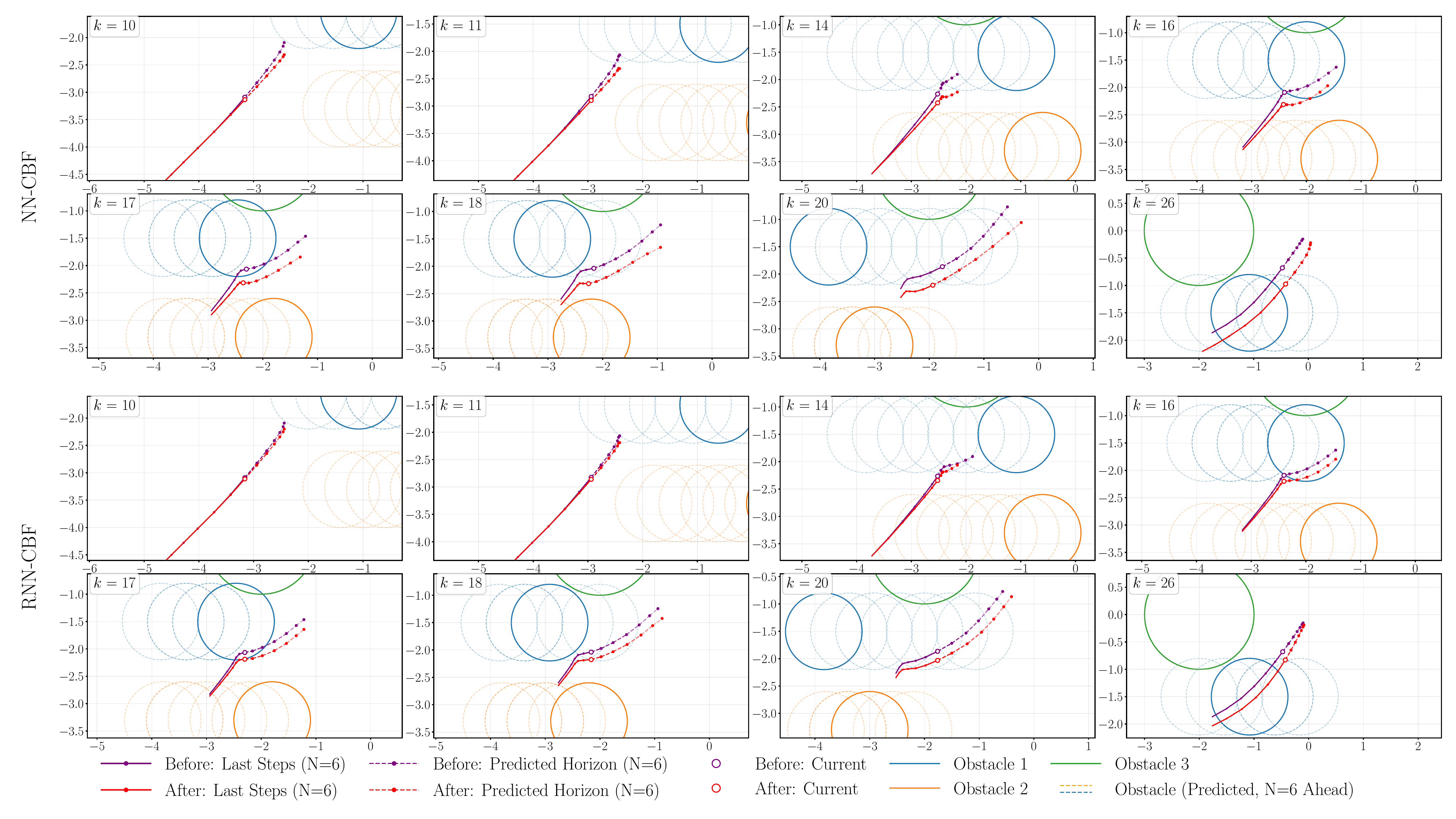}
  \caption{Snapshots of the NN-CBF and RNN-CBF policy, where `before' refers to before training and `after' refers to after training.}
  \label{fig:NN_RNN_policy_snapshots}
\end{figure*}

The RNN outputs follow the same trend as in the NN case. The $\gamma^\text{RNN}_{k,1}$ values for the first dynamic obstacle decrease between $k=8$ and $k=15$, but less sharply than in the NN, leaving a higher decay rate and producing less conservative behavior, as seen at $k=16$ in Fig. \ref{fig:NN_RNN_policy_snapshots}. Beyond this, the key distinction lies in training efficiency. The RNN learns the same policy with roughly 170 fewer episodes \citep{Dzhumageldyev2025}, highlighting the benefit of its recurrent structure. While further tests are needed to confirm, these results suggest that the RNN-based policy can potentially learn faster than the NN counterpart. 

\section{Conclusion} \label{sec:conclusion}

In this paper we proposed novel methods for merging MPC, RL and CBFs. The central idea is to use a parameterized MPC scheme, with learnable CBF constraints, as a function approximator to learn a safer and more performing policy. The proposed methods offer different ways of parameterizing the corresponding optimization problem and the class $\mathcal{K}$ function in the CBF condition, including neural network-based solutions. Numerical experiments showcase the ability of the proposed methods in learning safe and performing policies in two different obstacle avoidance tasks of increasing complexity. Further research will focus on expanding the current work to other RL algorithms (e.g., policy gradient methods), as well as investigating the case in which the true system dynamics are not known and how safety can still be enhanced via approximate CBFs during RL training.

\bibliography{ifacconf} 
                                                   
\end{document}